*Research Article*

# Surface and Buildup Region Dose Measurements with Markus Parallel-Plate Ionization Chamber, GafChromic EBT3 Film, and MOSFET Detector for High-Energy Photon Beams

**Ugur Akbas, Nazmiye Donmez Kesen, Canan Koksal, and Hatice Bilge**

*Division of Medical Physics, Istanbul University Oncology Institute, Istanbul, Turkey*

Correspondence should be addressed to Ugur Akbas; u.akbas@yahoo.com

The aim of the study was to investigate surface and buildup region doses for 6 MV and 15 MV photon beams using a Markus parallel-plate ionization chamber, GafChromic EBT3 film, and MOSFET detector for different field sizes and beam angles. The measurements were made in a water equivalent solid phantom at the surface and in the buildup region of the 6 MV and 15 MV photon beams at 100 cm source-detector distance for $5 \times 5$, $10 \times 10$, and $20 \times 20$ cm$^2$ field sizes and 0°, 30°, 60°, and 80° beam angles. The surface doses using 6 MV photon beams for $10 \times 10$ cm$^2$ field size were found to be 20.3%, 18.8%, and 25.5% for Markus chamber, EBT3 film, and MOSFET detector, respectively. The surface doses using 15 MV photon beams for $10 \times 10$ cm$^2$ field size were found to be 14.9%, 13.4%, and 16.4% for Markus chamber, EBT3 film, and MOSFET detector, respectively. The surface dose increased with field size for all dosimeters. As the angle of the incident radiation beam became more oblique, the surface dose increased. The effective measurement depths of dosimeters vary; thus, the results of the measurements could be different. This issue can lead to mistakes at surface and buildup dosimetry and must be taken into account.

## 1. Introduction

The deposited dose at the boundary between phantom and air is defined as the surface dose. Surface dose in radiation therapy is important in cases where the patient skin is dose-limiting tissue or part of the target volume in the treatment area. Acute skin reactions or delayed effects may occur after an overexposure of the skin. A therapeutic photon beam has electron contamination in the first few millimeters of skin caused by photon interactions in air or interactions with collimator and such scattering materials in the path of the beam. The surface dose depends on field size, source to skin distance (SSD), beam angle, beam energy, and beam modifiers such as blocks and multileaf collimator (MLC) systems. Accurate knowledge of surface dose is important, but the measurement of the dose at such shallow depths is a challenging issue. Due to each dosimetric tool having its own specific physical property, the results of surface dose measurements may vary.

Another challenging issue is defining the correct measurement depth for the proper measurement device. Effective measurement depths vary for each dosimeter. The International Commission on Radiation Units and Measurements (ICRU) and the International Commission on Radiological Protection (ICRP) recommend a 0.07 mm depth that generally corresponds to the interface between the dermis and epidermis layers of the skin for skin dose assessment [1]. Charged particle equilibrium does not exist at this depth and the dose gradient is high in the buildup region. Therefore, the choice of a suitable measurement device is important.

Surface and buildup region doses are measured with extrapolation chambers most accurately, but not every institution has this equipment. Parallel-plate ionization chambers are only good alternatives to extrapolation chambers due to their thin entrance window, but these chambers over-responded while measuring in the buildup region, based upon their internal dimensions. The overresponse occurs by secondary electrons scattering from the sidewall of the chamber. Gerbi's overresponse correction factors can be used for all types of fixed parallel-plate chambers. These factors are specific to chamber properties, volume, plate separation, and guard size. Because of their size and physical geometry, parallel-plate chambers are only suitable for phantom measurements [2, 3].

Radiochromic film is a substantial dosimeter for surface dosimetry, which has alleviated some problems faced with conventional radiation dosimetry. Radiochromic films have high spatial resolution and low spectral sensitivity. These characteristics make the films suitable for the measurement in regions of steep dose gradient [4, 5]. Devic et al. [1] presented correction factors for radiochromic films assuming the skin depth is 0.07 mm, by considering the effective depth of measurement. Bilge et al. [6, 7] used EBT2 film in their surface dose study and compared the results to those from a parallel-plate chamber. They found the difference between EBT film and ionization chamber to be within 5% for 6 MV and 3% for 18 MV.

Metal oxide semiconductor field effect transistor (MOSFET) dosimeters, belonging to the category of the semiconductor detectors, have been used for surface dose measurement. MOSFET dosimeters have several advantages over the conventional dosimeters including their small physical size, immediate readout and reuse, ability of multiple point dose measurement, and ability of recording dose history. MOSFET dosimeters have been presented as a user-friendly and effective alternative to radiochromic film and thermoluminescent dosimeters (TLD) for surface measurements [8]. The data with standard-size MOSFET shows large differences of up to 50% between MOSFET and TLD or film measurement at the surface, which is relevant to the different effective buildup thicknesses of these dosimeters [9]. As a consequence, the buildup thickness of all dosimeters should be taken into consideration for an accurate surface dose measurement.

The ion beam penetration range in a material is often characterized by the water equivalent thickness (WET). WET measures the thickness of liquid water needed to stop the ion beam in the same manner that a certain thickness of the given material might [10]. Dosimeters are made from different materials and each of them has its own effective depth of measurement. Thus, for a decent surface dose measurement, WET of dosimeters should be taken into account.

The aim of this study is to investigate surface doses using a parallel-plate ionization chamber, EBT3 films, and MOSFET dosimeters in different field sizes and beam angles for 6 MV and 15 MV high-energy photon beams.

## 2. Materials and Methods

*2.1. Phantom and Irradiation Conditions.* The percentage depth dose (PDD) at the central axis in the buildup region was measured in a water equivalent RW3 slab phantom (SP34, PTW Freiburg, Freiburg, Germany). The solid phantom has negligible uncertainty compared to liquid water, and also helps to reduce the uncertainties in depths [11]. The physical density of RW3 water equivalent phantom is 1.045 g cm$^{-3}$ and the phantoms consist of $40 \times 40$ cm$^2$ slabs of various thicknesses ranging from 1–10 mm. The equivalent depth was referred to the equivalent mass depth from the phantom surface to the effective measurement point of the dosimeter by considering the density of the RW3 water equivalent phantom.

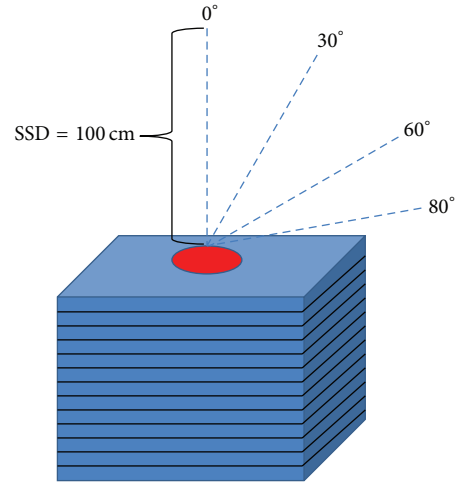

Figure 1: Schematic of oblique beam irradiation with gantry angles.

The surface dose measurements were carried out with 3 different dosimeters, including the Markus parallel-plate ionization chamber (Markus 23343, PTW Freiburg, Freiburg, Germany), EBT3 GafChromic® film (International Specialty Product, NJ, US), and a commercial MOSFET (TN-502, Thomson and Nielson Ltd., Ottawa, Canada). These 3 dosimeters were irradiated for $5 \times 5$, $10 \times 10$, and $20 \times 20$ cm$^2$ field sizes at 0, 1, 2, 5, 10, and 15 mm phantom depths that were assumed to be buildup region in a source-detector distance (SDD) setup with Varian Trilogy linear accelerator (Varian, Palo Alto, CA). Source to skin distance correction for SSD = 100 cm was applied to the results. In this study, effect of the oblique incidence beam was also investigated at 0°, 30°, 60°, and 80° beam angles (see Figure 1). 6 MV and 15 MV photon beams were used during irradiation and the measurements were repeated 3 times to acquire an average value.

*2.2. Parallel-Plate Ionization Chamber Measurements.* A Markus parallel-plate ionization chamber was used for the surface dose measurement. In this study, the surface depth was assumed to be 0.07 mm. The physical effective point of measurement for Markus chamber was defined as 0.023 mm, at the inner surface of the proximal collecting plate. The plate separation is fixed at 2 mm; the sidewall-to-collector distance is 0.35 mm. The relative ionization for the points of interest was acquired by dividing the charge collected at depth by the charge at the buildup depth using the Unidose dosimeter (PTW Freiburg, Germany). The polarity effect of the chamber was considered and the readings were corrected with the following formula:

$$Q_{\text{avg}} = \frac{(Q_+ + Q_-)}{2}. \tag{1}$$

$Q_{avg}$ is the average charge used for relative ionization where $Q_+$ and $Q_-$ are the charges accumulated with positive and negative polarities, respectively.

Surface and buildup region doses were measured at 100 cm fixed SDD, and SSD correction was made for SSD = 100 cm. $40 \times 40 \times 10$ cm$^3$ size solid water phantom was formed for the measurements. Measurements were carried out for 6 MV and 15 MV photon beams by delivering 100 MU and the results were normalized to 15 mm and 30 mm depths of phantom as the accepted maximum dose depth, respectively. To evaluate the response of other dosimeters to oblique incident beams, the Markus chamber was taken as a reference and PDDs were measured at the surface of phantom for beam angles of 0°, 30°, 60°, and 80° by using a $10 \times 10$ cm$^2$ field size.

Gerbi's method was applied to PDDs obtained from the chamber for correcting the overdoses in the buildup region [11]. Consider

$$P'(d, E) = P(d, E) - \xi'(0, E) l e^{-\alpha(d/d_{max})},$$
$$\xi(0, E) = [-1.666 + (1.982 \text{IR})] \quad (2)$$
$$\times (C - 15.8) \ (\%/\text{mm}).$$

$\xi(0, E)$ = energy dependent chamber factor that indicates the overresponse per mm of chamber plate separation at the surface of the phantom. The values −1.666, 1.982, and 15.8 are constants that were taken from the graph, which represent the % maximum ionization per mm of plate separation at the phantom surface plotted as a function of guard width or collector edge-sidewall distance [12],

IR = ionization ratio at depths of 20 cm and 10 cm, which is measured at a fixed source-detector distance and $10 \times 10$ cm$^2$ field size. IR values are 0.672 and 0.763 for 6 MV and 15 MV photon beams, respectively. These values of IR is measured and compared to [13],

$P'$ = corrected percent depth dose,

$P$ = relative depth ionization,

$E$ = energy,

$d_{max}$ = maximum dose depth,

$C$ = sidewall-collector distance (0.35 mm for PTW-Markus 23343),

$l$ = plate separation (2 mm for PTW-Markus 23343),

$\alpha$ = 5.5, constant,

$d$ = depth of the chamber front window ($d = 0$ for surface) [12],

$\xi'(0, E) l e^{-\alpha(d/d_{max})}$ = calculated correction factors used for 6 MV and 15 MV photon beams at various depths.

*2.3. Film Measurements.* In this study, GafChromic EBT3 film (International Specialty Product, NJ, US) was used. The EBT3 film has a single active layer of approximately 30 μm thickness. This active layer is sandwiched between two 125 μm thick transparent polyester sheets. Compared to EBT2 film, the EBT3 film has a symmetric structure that allows for scanning either side. Matte polyester substrate was used for EBT3 film instead of smooth polyester substrate to prevent Newton's rings formation. Also the EBT3 film is more sensitive than the older versions of radiochromic film with its wide dose range from 1 cGy to 40 Gy. The effective point of measurement for the EBT3 film was defined at 0.153 mm depth [1, 14].

For the investigation of surface dose, EBT3 film sheets were used from batch number A03051204. A calibration curve was created before irradiation to make evaluation accurate. The EBT3 film was cut into small pieces of $2.5 \times 2.5$ cm$^2$ size and placed perpendicularly between the solid water slab phantoms at the depth of 5 cm where 1 cGy equals to 1 MU. The EBT3 films were irradiated with doses ranging from 0–800 cGy at $10 \times 10$ cm$^2$ field size. Unirradiated (0 cGy) film was used as a background. After a period of 24 hours after irradiation, the exposed films were scanned and digitized using an Epson Expression 10000XL scanner (Epson America, Long Beach, CA, USA). After the scanning process, each film was separated into blue, green, and red channels using ImageJ software and the red channel was chosen for its high contrast. The average readings of optical densities (OD) of film pieces were acquired using a commercial software (PTW Mephysto mc$^2$, PTW Freiburg, Freiburg, Germany). The average background OD was subtracted from each irradiated film piece to obtain net OD. Then, these net ODs were corrected to known dose values to create the calibration curve. The calibration curve was used for converting the net ODs to the dose.

The film measurements for surface dose were made under the same setup condition, field sizes, and beam angles of parallel-plate chamber measurements.

*2.4. MOSFET Measurements.* The surface dose measurements were carried out using TN502RD mobile MOSFETs (Best Medical, Canada) of standard sensitivity. The MOSFET dosimeter is an electronic device. The physical dimensions of the dosimeter are 2.5 mm long, 2 mm wide, and 0.3 mm thick. It measures integrated dose with its silicon chip of 1 mm$^2$ size. It has an active area of 0.2 mm$^2$ covered by an epoxy bulb. The MOSFET detector has an intrinsic buildup equal to a WET of 0.8 mm for epoxy side and 1.8 mm for flat side of the detector (see Figure 2) [15]. The dose verification system consists of a TN RD 70 W reader module, dose verification software, and a wireless transceiver. The reader module has dual bias settings; 1 mV/cGy for standard bias setting and 2.7 mV/cGy for high sensitivity bias setting. The standard bias setting was used in this study.

The MOSFET dosimeters were calibrated before the first measurement to obtain the maximum measurement accuracy. The dosimeters were placed under 6 MV and 15 MV photon beams at a specified dose level where 1 cGy equals 1 MU, and then the output voltage of the dosimeters was compared to the set level. The measured voltage value to the actual value of the radiation dose ratio gave the calibration factors of each dosimeter. For the calibration setup, the water equivalent slab phantoms and a special acrylic phantom size of $30 \times 30 \times 10\, cm^3$ were used. The acrylic phantom has 5 hollows at the center of the surface to place the dosimeters. In surface and buildup region dose measurements, the water equivalent phantoms were added up to create physical depths; thus, the flat side of the MOSFET dosimeter was used while the epoxy side was placed into these hollows to avoid air gap. The irradiation for calibration was made with 5 dosimeters for $10 \times 10\, cm^2$ field size at 5 cm and 10 cm phantom depths for 6 MV and 15 MV, respectively. The changes in the threshold voltage $V_{TH}$ were recorded and the calibration factors were obtained using the software. The dosimeter that had the smallest standard deviation value was chosen for the measurements.

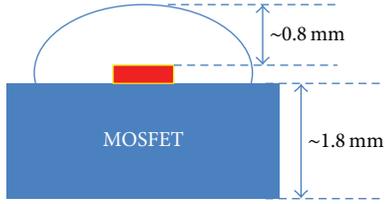

Figure 2: Schematic of typical MOSFET with conventional epoxy bubble.

The surface measurement was made using the calibrated dosimeter. For the comparison, the same field sizes, depths, and MU values of the both film and parallel-plate chamber measurements were used. The measurements of obliquely incident beams were taken for the same angles with Markus chamber and film to evaluate the response of MOSFET dosimeters. In addition, the epoxy side of the MOSFET dosimeter was used to investigate the angular dependence for in vivo usage.

## 3. Results

For 6 MV and 15 MV, the percentage doses at the surface and buildup regions for different field sizes measured using the Markus parallel-plate ionization chamber, EBT3 film, and MOSFET dosimeter are given in Tables 1, 2, and 3. In this study, PDD values were investigated at the depth of 0.07 mm, which was suggested for surface dose measurements by ICRU Report 39 [16]. To obtain the dose at 0.07 mm depth, an interpolation calculation was made by comparing the PDD data to a four-order polynomial fit of Markus chamber measurements in the buildup region. For EBT3 film and MOSFET dosimeter measurements, extrapolation was made because their effective depths of measurement were just above the phantom.

The dosimeters have their specific effective depth of measurement. For this reason, a measurement point at a phantom depth is not the same for all dosimeters. The WET must be considered to make an accurate comparison. Thus, the PDD values in Tables 1, 2, and 3 were used to calculate the doses at the same effective depths in the buildup region.

Table 1: (a) Percentage depth doses (PDDs) obtained with a Markus chamber in a water equivalent RW3 phantom using 6 MV photon beams for different field sizes at SSD = 100 cm. (b) Percentage depth doses (PDDs) obtained with a Markus chamber in a water equivalent RW3 phantom using 15 MV photon beams for different field sizes at SSD = 100 cm.

(a)

| Phantom depth (mm) | WET (mm) | PDD (%) | | |
|---|---|---|---|---|
| | | $5 \times 5\, cm^2$ | $10 \times 10\, cm^2$ | $20 \times 20\, cm^2$ |
| 0 | 0.023 | 10.8 ± 0.8 | 16.6 ± 0.9 | 28.1 ± 0.8 |
| 0.067 | 0.07 | 14.6 ± 0.9 | 20.3 ± 0.9 | 31.4 ± 0.8 |
| 1 | 1.068 | 38.7 ± 0.8 | 43.4 ± 0.8 | 52.5 ± 0.7 |
| 2 | 2.113 | 57.1 ± 0.7 | 61.0 ± 0.6 | 68.2 ± 0.6 |
| 5 | 5.248 | 83.6 ± 0.6 | 85.9 ± 0.4 | 90.0 ± 0.4 |
| 10 | 10.473 | 97.8 ± 0.2 | 98.4 ± 0.2 | 99.2 ± 0.3 |
| 15 | 15.698 | 100.0 ± 0.1 | 100.0 ± 0.1 | 100.0 ± 0.2 |

(b)

| Phantom depth (mm) | WET (mm) | PDD (%) | | |
|---|---|---|---|---|
| | | $5 \times 5\, cm^2$ | $10 \times 10\, cm^2$ | $20 \times 20\, cm^2$ |
| 0 | 0.023 | 6.9 ± 0.9 | 13.7 ± 0.8 | 26.7 ± 0.8 |
| 0.067 | 0.07 | 8.0 ± 0.9 | 14.9 ± 0.8 | 27.9 ± 0.8 |
| 1 | 1.068 | 22.7 ± 0.8 | 29.6 ± 0.7 | 42.0 ± 0.8 |
| 2 | 2.113 | 34.8 ± 0.7 | 41.5 ± 0.6 | 53.5 ± 0.9 |
| 5 | 5.248 | 60.3 ± 0.8 | 65.8 ± 0.7 | 75.6 ± 0.8 |
| 10 | 10.473 | 81.3 ± 0.6 | 85.4 ± 0.5 | 92.1 ± 0.6 |
| 15 | 15.698 | 92.3 ± 0.4 | 94.8 ± 0.3 | 98.6 ± 0.4 |
| 20 | 20.923 | 97.1 ± 0.3 | 98.7 ± 0.3 | 101.0 ± 0.9 |
| 25 | 26.148 | 99.6 ± 0.3 | 100.1 ± 0.3 | 100.9 ± 0.7 |
| 30 | 31.373 | 100.0 ± 0.2 | 100.0 ± 0.2 | 100.0 ± 0.3 |

Interpolation was made for the first 5 mm of the phantom for each dosimeter. The results for $5 \times 5$, $10 \times 10$, and $20 \times 20\, cm^2$ field sizes are shown in Figure 3 for 6 MV and Figure 4 for 15 MV.

In this study, the responses of dosimeters for the changing field size were investigated. Surface doses at 0.07 mm were normalized as a percentage of the maximum dose for $5 \times 5$, $10 \times 10$, and $20 \times 20\, cm^2$ fields at 100 cm fixed SSD. As can be seen from Figure 5, the percentage dose at the surface increased with field size for each dosimeter, as expected.

To assess the angular response of each dosimeter at the surface, the PDD was measured at the slab phantom surface with oblique beam angles of 30°, 60°, and 80°, using 100 cm fixed SSD and $10 \times 10\, cm^2$ field size. To normalize the PDDs, the measurements were made at dose maximum depths with same field size and gantry angle set to 0° for each dosimeter. The results are presented in Figure 6 for 6 MV and 15 MV.

The epoxy side of the MOSFET dosimeter was also used to evaluate the angular dependence for in vivo usage. The same SSD, field size, and beam angles were used. The results were normalized to 0° beam angle result. The results are shown in Figure 7 for 6 MV and 15 MV.

Table 2: (a) Percentage depth doses (PDDs) obtained with EBT3 films in a water equivalent RW3 phantom using 6 MV photon beams for different field sizes at SSD = 100 cm. (b) Percentage depth doses (PDDs) obtained with EBT3 films in a water equivalent RW3 phantom using 15 MV photon beams for different field sizes at SSD = 100 cm.

(a)

| Phantom depth (mm) | WET (mm) | PDD (%) | | |
|---|---|---|---|---|
| | | $5 \times 5$ cm$^2$ | $10 \times 10$ cm$^2$ | $20 \times 20$ cm$^2$ |
| N/A | 0.07 | 12.8 ± 1.7 | 18.8 ± 1.8 | 27.1 ± 1.8 |
| 0 | 0.153 | 15.4 ± 1.9 | 20.4 ± 1.8 | 30.5 ± 1.6 |
| 1 | 1.198 | 35.4 ± 1.8 | 48.0 ± 1.7 | 55.3 ± 1.5 |
| 2 | 2.243 | 55.5 ± 1.8 | 61.3 ± 1.8 | 75.2 ± 1.6 |
| 5 | 5.378 | 82.4 ± 1.8 | 87.6 ± 0.9 | 85.2 ± 0.9 |
| 10 | 10.603 | 96.4 ± 1.6 | 99.2 ± 0.9 | 97.4 ± 1.2 |
| 15 | 15.828 | 100.0 ± 1.4 | 100.0 ± 1.1 | 100.0 ± 1.0 |

(b)

| Phantom depth (mm) | WET (mm) | PDD (%) | | |
|---|---|---|---|---|
| | | $5 \times 5$ cm$^2$ | $10 \times 10$ cm$^2$ | $20 \times 20$ cm$^2$ |
| N/A | 0.07 | 7.3 ± 1.4 | 13.4 ± 1.8 | 26.8 ± 1.3 |
| 0 | 0.153 | 10.6 ± 1.6 | 14.1 ± 1.5 | 28.0 ± 1.4 |
| 1 | 1.198 | 20.0 ± 1.3 | 31.8 ± 1.4 | 38.4 ± 1.2 |
| 2 | 2.243 | 36.2 ± 1.2 | 40.8 ± 1.6 | 51.2 ± 1.6 |
| 5 | 5.378 | 59.1 ± 1.5 | 61.8 ± 1.4 | 73.4 ± 1.5 |
| 10 | 10.603 | 75.4 ± 1.4 | 76.4 ± 1.6 | 92.3 ± 1.2 |
| 15 | 15.828 | 87.4 ± 0.9 | 86.1 ± 1.5 | 104.4 ± 1.9 |
| 20 | 21.053 | 94.8 ± 1.2 | 91.9 ± 1.4 | 105.8 ± 1.9 |
| 25 | 26.278 | 97.8 ± 1.0 | 95.9 ± 1.1 | 104.3 ± 1.8 |
| 30 | 31.503 | 100.0 ± 1.0 | 100.0 ± 1.1 | 100.0 ± 1.8 |

Table 3: (a) Percentage depth doses (PDDs) obtained with a MOSFET dosimeter in a water equivalent RW3 phantom using 6 MV photon beams for different field sizes at SSD = 100 cm. (b) Percentage depth doses (PDDs) obtained with a MOSFET dosimeter in a water equivalent RW3 phantom using 15 MV photon beams for different field sizes at SSD = 100 cm.

(a)

| Phantom depth (mm) | WET (mm) | PDD (%) | | |
|---|---|---|---|---|
| | | $5 \times 5$ cm$^2$ | $10 \times 10$ cm$^2$ | $20 \times 20$ cm$^2$ |
| N/A | 0.07 | 17.4 ± 1.3 | 25.5 ± 1.3 | 34.5 ± 1.8 |
| 0 | 1.800 | 40.1 ± 1.4 | 46.4 ± 1.6 | 57.2 ± 1.6 |
| 1 | 2.845 | 56.8 ± 1.0 | 60.6 ± 1.5 | 69.3 ± 1.5 |
| 2 | 3.890 | 66.5 ± 1.2 | 70.2 ± 1.6 | 77.2 ± 1.6 |
| 5 | 7.045 | 87.8 ± 1.1 | 89.0 ± 1.4 | 97.5 ± 1.7 |
| 10 | 12.250 | 106.3 ± 0.8 | 105.2 ± 1.2 | 105.5 ± 1.9 |
| 15 | 17.475 | 100.0 ± 0.9 | 100.0 ± 1.1 | 100.0 ± 1.8 |

(b)

| Phantom depth (mm) | WET (mm) | PDD (%) | | |
|---|---|---|---|---|
| | | $5 \times 5$ cm$^2$ | $10 \times 10$ cm$^2$ | $20 \times 20$ cm$^2$ |
| N/A | 0.07 | 9.0 ± 1.6 | 16.4 ± 1.9 | 28.9 ± 1.8 |
| 0 | 1.800 | 24.6 ± 1.7 | 33.0 ± 1.9 | 40.0 ± 1.7 |
| 1 | 2.845 | 30.0 ± 1.6 | 40.0 ± 1.8 | 52.4 ± 1.6 |
| 2 | 3.890 | 42.3 ± 1.5 | 50.3 ± 1.6 | 59.1 ± 1.6 |
| 5 | 7.045 | 62.1 ± 1.6 | 70.0 ± 1.7 | 74.8 ± 1.4 |
| 10 | 12.250 | 79.4 ± 1.7 | 88.7 ± 1.6 | 90.8 ± 1.5 |
| 15 | 17.475 | 89.9 ± 1.3 | 97.9 ± 1.5 | 95.2 ± 1.6 |
| 20 | 22.700 | 96.7 ± 1.2 | 99.0 ± 1.6 | 99.2 ± 1.8 |
| 25 | 27.925 | 98.3 ± 1.3 | 99.3 ± 1.7 | 99.8 ± 1.8 |
| 30 | 33.150 | 100.0 ± 1.0 | 100.0 ± 1.2 | 100.0 ± 1.1 |

## 4. Discussion

Surface dose measurement is one of the most challenging issues for clinical dosimetry in radiotherapy. Accurate knowledge of surface and buildup region doses is important. In clinical radiotherapy, different field sizes can be used with different gantry angles. These irradiation conditions affect the doses in the buildup region. The choice of an appropriate tool for surface dosimetry is necessary. Due to their physical characteristics, the response of dosimetric tools may vary for superficial doses. In this study, the surface dose was measured with a parallel-plate ion chamber, EBT3 film, and MOSFET detectors for different field sizes, different depths, and oblique beams on a phantom surface to obtain the dose response of these dosimetric tools. Also, the availability of in vivo usage was investigated.

Surface and buildup region doses are best measured with extrapolation chambers [17]. Parallel-plate ionization chambers can be utilized after having used Gerbi's overresponse correction factors. In this study, the Markus parallel-plate ionization chamber was used as a reference to evaluate the results. Surface doses with the Markus chamber were found to be 10.8%, 16.6%, and 28.1% at 0 mm for $5 \times 5$, $10 \times 10$, and $20 \times 20$ cm$^2$ field sizes, respectively. Bilge et al. [6] investigated the surface dose for 6 and 18 MV (Siemens Oncor Impression Plus linear accelerator) photon beams using a Markus parallel-plate ion chamber and EBT2 film. The results with Markus chamber for 6 MV were found to be 10.0%, 15.0%, 23.0%, and 35.0% ± 1% (1 SD) at 0 mm for $5 \times 5$, $10 \times 10$, $20 \times 20$, and $30 \times 30$ cm$^2$ field sizes, respectively. Yadav et al. [18] reported a technical note about skin dose estimation. In their study, surface dose measurements were carried out for 6 MV (Elekta precise linear accelerator) photons, for various field sizes, with beam modifiers at different SSDs. The result for the open $10 \times 10$ cm$^2$ field was 14.8% at 100 cm SSD.

Jong et al. [19] measured the surface dose with Markus chamber for 6 and 10 MV (Varian Clinac 2100 C/D accelerator) photons at the depth of 0 mm and the results were found to be 15.8 ± 0.03% and 11.8 ± 0.00%, respectively. Qi et al. [20] obtained PDDs with an Attix chamber in water equivalent phantom for 6 MV (Varian 600C linear accelerator) photons. At fixed 100 cm SSD, for $5 \times 5$, $10 \times 10$, $20 \times 20$, and $30 \times 30$ cm$^2$ field sizes, surface doses were found to be 12.9%, 18.9%, 29.1%, and 37.9%, respectively. They found these results using 0.048 mm WET which equals to 0 mm phantom depth. In our study, we interpolated the obtained data to acquire the surface dose at 0.07 mm WET (14.6%, 20.3%, and 31.4% for $5 \times 5$, $10 \times 10$, and $20 \times 20\,cm^2$ field sizes, resp.).

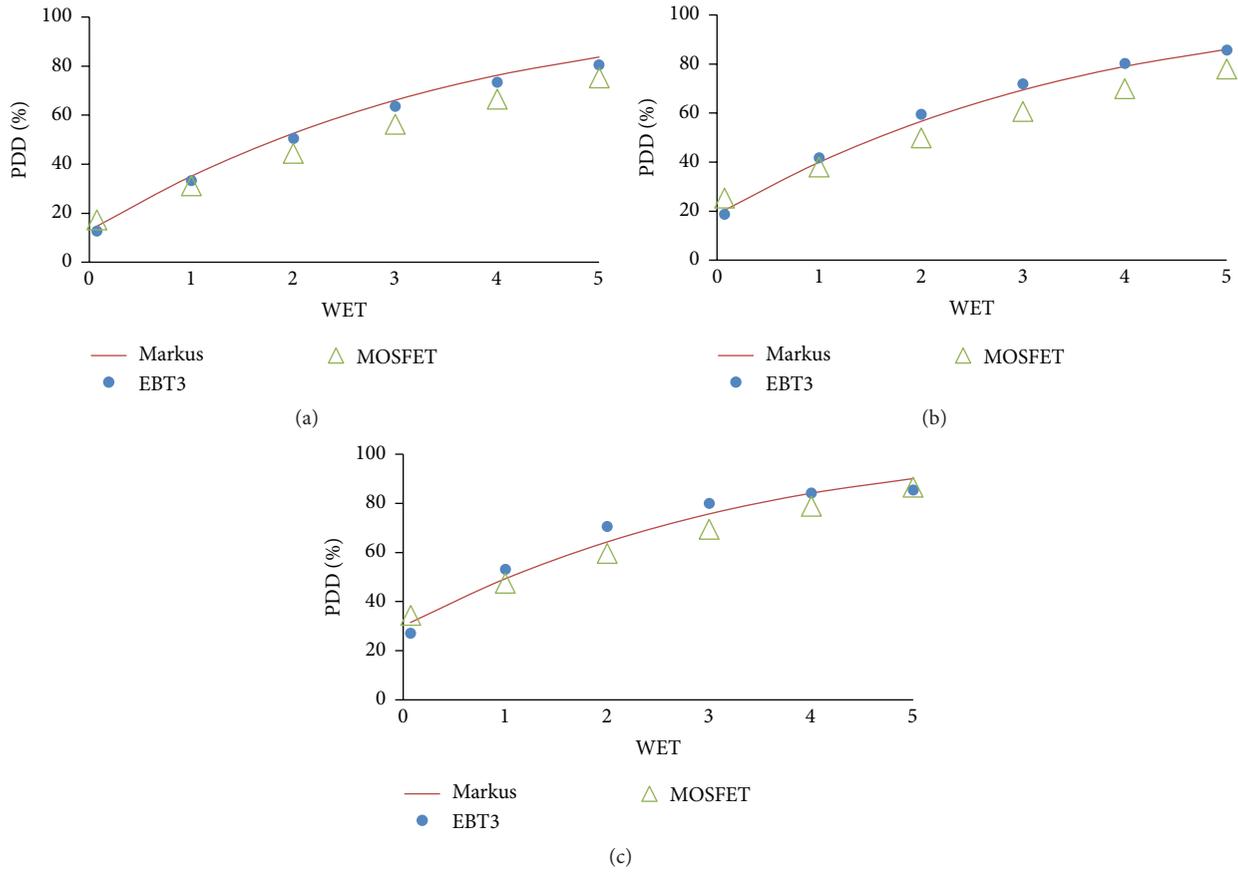

Figure 3: Percentage depth doses (PDDs) for normally incident 6 MV photon beams at open fields with different sizes at 100 cm fixed SSD. The doses at 0, 1, 2, 3, 4, and 5 mm effective depths for each dosimeter were calculated by interpolation, and the following field sizes were investigated: $5 \times 5\,cm^2$ (a), $10 \times 10\,cm^2$ (b), and $20 \times 20\,cm^2$ (c).

Radiochromic EBT films are good alternative detectors for surface dosimetry with their high spatial resolution and low spectral sensitivity [12, 21]. The results with EBT3 films were found to be 15.4%, 20.4%, and 30.5% at the depth of 0 mm for $5 \times 5$, $10 \times 10$, and $20 \times 20\,cm^2$ field sizes, respectively. The EBT3 film has a WET of 0.153 mm which equals 0 mm phantom depth. The measured buildup region doses were used for extrapolation calculation to acquire the dose at 0.07 mm WET. The results were 12.8%, 18.8%, and 27.1% for the same field sizes at 0.07 mm. Devic et al. [1] investigated skin dose with radiochromic EBT model film at 0.07 and 0.153 mm depths for 6 MV (Varian Clinac 2100 C/D linear accelerator) photons. They found the surface doses 11.5%, 17.0%, and 28.4% at 0.07 mm for $5 \times 5$, $10 \times 10$, and $20 \times 20\,cm^2$ field sizes, respectively. At 0.153 mm depth, they measured the surface doses 14.3%, 19.9%, and 28.4% for the same field sizes, respectively. Qi et al. [20] also searched for surface dose with an EBT film at the phantom surface where the WET value equals 0.153 mm; the results were 13.5%, 23.5%, and 33.7% for the same field sizes stated above, respectively. Bilge et al. [6] used EBT film and measured the surface dose to be $15.0 \pm 2\%$, $20.0 \pm 2\%$, $28.0 \pm 2\%$, and $40.0 \pm 2\%$ at 0 mm depth for $5 \times 5$, $10 \times 10$, $20 \times 20$, and $30 \times 30\,cm^2$ field sizes, respectively. Our EBT3 film results for surface dose measurements are in close agreement with our Markus results. Also, our study shows coherency with other publications.

MOSFET is a promising dosimeter with advantages of immediate response, small size, good reproducibility, and being independent on dose rate for in vivo surface dose measurements [22]. In this study, the flat side of the MOSFET detector, which has a WET of 1.8 mm, was used for surface dose measurements. The surface doses at 0 mm depth were found to be 40.1%, 46.4%, and 57.2% for $5 \times 5$, $10 \times 10$, and $20 \times 20\,cm^2$ field sizes, respectively. These results present the doses at 1.8 mm WET. The extrapolation calculation was applied to the MOSFET results and surface doses at 0.07 mm were acquired to be 17.4%, 25.5%, and 34.5% for the same field sizes, respectively. Qi et al. [20] measured the surface and buildup region doses with MOSkin, which is the newly designed miniature MOSFET detector. They found the surface doses at 0.00 mm WET to be $13.5 \pm 1.8\%$, $19.5 \pm 1.7\%$, and $30.1 \pm 1.7\%$ and they also measured the dose at 1.045 mm WET to be $40.6 \pm 1.6\%$, $46.0 \pm 1.5\%$, and $52.7 \pm 1.7\%$ for $5 \times 5$, $10 \times 10$, and $20 \times 20\,cm^2$ field sizes, respectively. Our results are in good agreement with the results of MOSkin when the WET values are considered. Elvan Erkan and Kemikler [23] examined the effect of thermoplastic masks used for immobilization on surface dose at Co-60 and 6 MV (Siemens Oncor Impression Plus linear accelerator) photon energies. They used MOSFET detectors which have a WET of 0.8 mm and their surface dose measurements were $32.2 \pm 2.3\%$, $37.4 \pm 2.3\%$, and $43.2 \pm 2.4\%$ for the same field sizes that we used, respectively. The WET values that were used in both studies obviously affect the results.

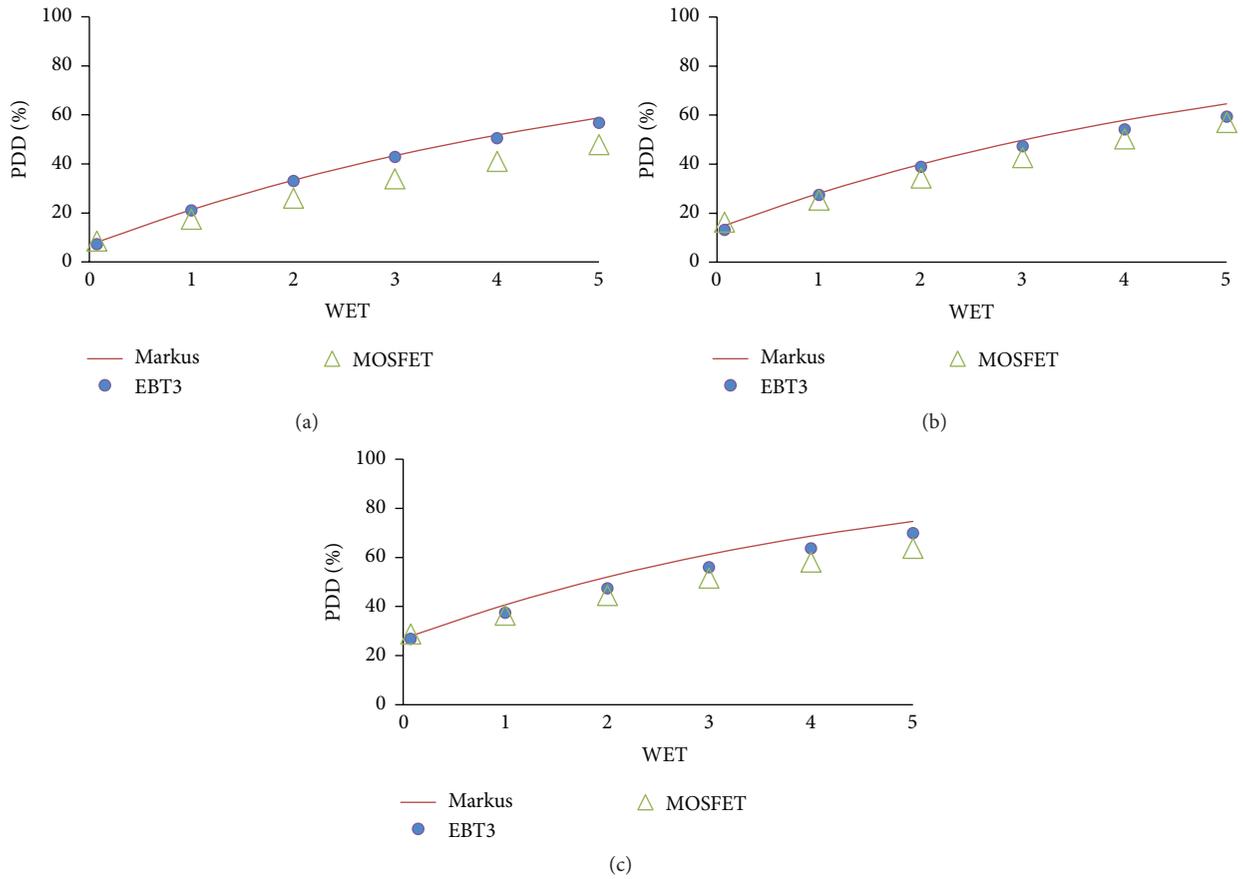

Figure 4: Percentage depth doses (PDDs) for normally incident 15 MV photon beams at open fields with different sizes at 100 cm fixed SSD. The doses at 0, 1, 2, 3, 4, and 5 mm effective depths for each dosimeter were calculated by interpolation, and the following field sizes were investigated: $5 \times 5$ cm$^2$ (a), $10 \times 10$ cm$^2$ (b), and $20 \times 20$ cm$^2$ (c).

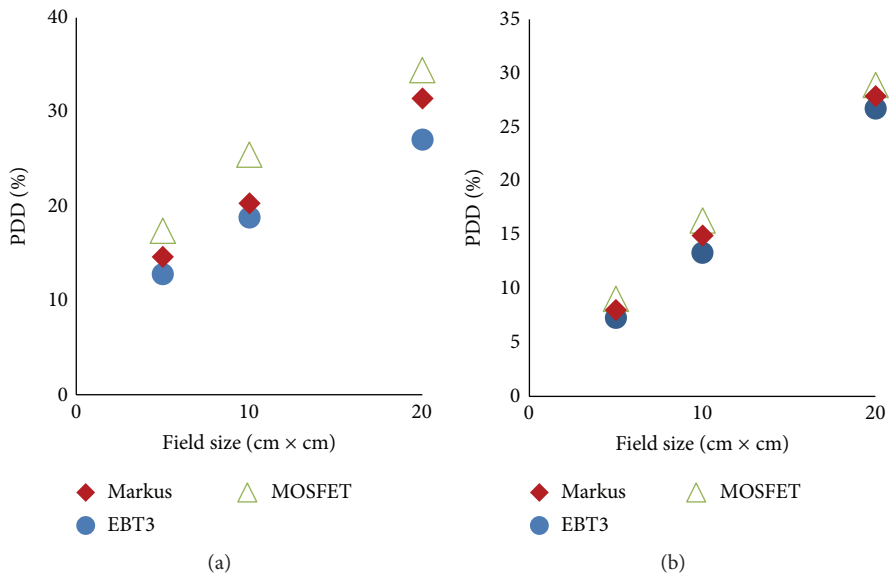

Figure 5: Percentage dose at the surface (0.07 mm) for 6 MV (a) and 15 MV (b) photon beams as a function of field size at 100 cm fixed SSD for Markus parallel-plate ionization chamber, EBT3 film, and MOSFET dosimeter.

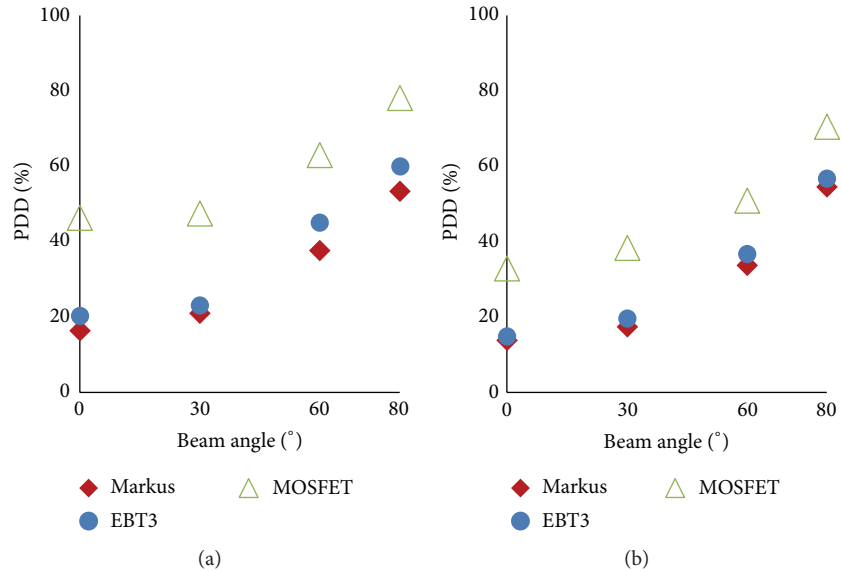

Figure 6: Percentage depth doses (PDDs) for 6 MV (a) and 15 MV (b) measured with Markus chamber, EBT3 film, and MOSFET detector, respectively, at the phantom surface for normal and oblique incident beams.

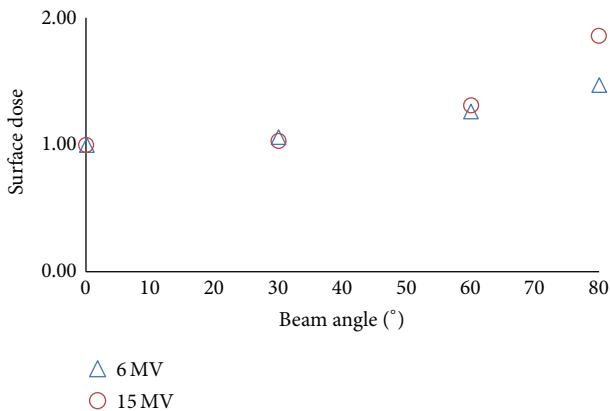

Figure 7: The angular dependence of the epoxy side of the MOSFET dosimeter for 6 MV and 15 MV.

The surface and buildup region doses were found to be the highest with MOSFET at the physical phantom depths. When the WET values of dosimeters are considered and the same water equivalent depths are taken into account, MOSFET results are slightly lower for the first 5 mm, as seen in Figure 3. The detectors that have deeper WET values increase the surface dose results if the surface dose depth is determined as 0.00 mm phantom depth. PDD values should be used for extrapolation or interpolation calculations to reach the doses at surface and buildup regions. Greater PDD values can increase the calculation accuracy, so the measurements should be performed from surface to maximum dose depth by 1 mm intervals if possible.

As expected, for all dosimeters, surface dose increases with field size due to extra electron contamination and photon head scatter (see Figure 5). The EBT3 film results for all field sizes and depths show closer agreement to the Markus chamber results.

As the angle of the incident radiation beam becomes more oblique, the surface dose increases and the $d_{max}$ moves towards the surface due to more secondary electrons being ejected along the oblique path of the beam [24]. The surface dose increases sharply at larger angles, with the relative dose being 50% larger at an obliquity of $\sim 55°$ [25, 26]. In this study, the effect of the oblique beams was investigated at 0 mm phantom depth. Figure 6 shows the results of the measured doses that have been normalized to the dose at $0°$ gantry angle. As the beam angle increases, the measured surface dose increases since the charged particle equilibrium (CPE) region shifts towards the surface. Maximum deviations were found at a beam incidence of $80°$ for all dosimeters. MOSFET detector measured the highest surface dose at all angles compared to the Markus chamber and EBT3 film. Scalchi et al. [27] demonstrated surface angular dependence on a special unencapsulated MOSFET dosimeter; and Qi et al. [20] measured the angular dependence at surface with Attix ionization chamber and MOSkin detector, finding that the oblique beam incidence increases the surface dose. Qin et al. [28] found the maximum deviation at an oblique angle of $72°$. The epoxy side of the MOSFET dosimeter showed less angular dependence at $30°$ and $60°$ while the result of $80°$ was still high. This decrease of the angular dependence at $30°$ and $60°$ makes MOSFET dosimeters more available for in vivo measurements. We believe that the major reason behind the differences of the surface dose deviation values between our study and these other publications is the different design of the dosimeters and irradiation conditions.

## 5. Conclusions

At the physical phantom depths, the surface and buildup region doses were found to be the highest with MOSFET. When the WET values of dosimeters are considered and the same water equivalent depths are taken into account, MOSFET results are slightly lower for the first 5 mm. The EBT3 film shows closer agreement to the Markus chamber at these depths as a consequence of WET of dosimeters. The dose differences between dosimeters are greater with 6 MV at all field sizes when the results for the surface dose measurements at 0.07 mm depth were compared. MOSFET gives much higher results compared to EBT3 and Markus at oblique measurements as a result of its WET value. This physical property of MOSFET causes a special consideration for in vivo usage.

The EBT3 film has advantages like its dosimetric properties of homogenous material and tissue equivalence, but it is not a practical dosimeter for in vivo usage. The EBT3 film requires a waiting time before scanning and a calibration curve to acquire the absolute doses. The MOSFET dosimeter is easy to use and gives immediate results. The MOSFET dosimeter gives a higher response at surface measurement but the overestimation of detectors can be measured with another reference dosimetry system. When a factor of overestimation is found, the MOSFET dosimeters can be used for in vivo surface dosimetry.

There are some publications about the surface dose measurements which take part in literature, but in these publications the surface doses were investigated at 0 mm physical depth. It is necessary to know the effective depth of the dosimeters. In this study, surface and buildup region doses were measured and calculated by considering the WET of the dosimeters.

The surface and buildup region doses should be evaluated considering the depth. The effective measurement depths of dosimeters vary; thus, the results of the measurements could be different. This issue can lead to mistakes at surface and buildup dosimetry and must be taken into account.

## Competing Interests

The authors declare that they have no competing interests.

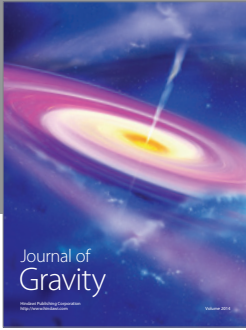
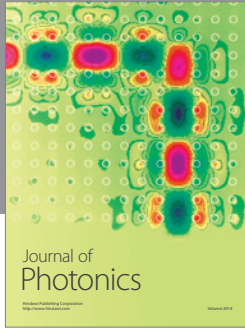
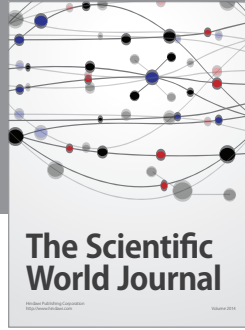
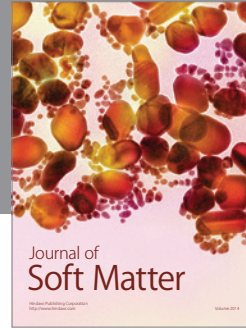
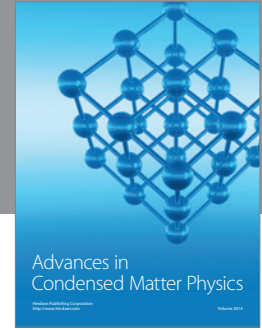
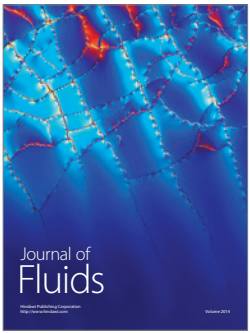
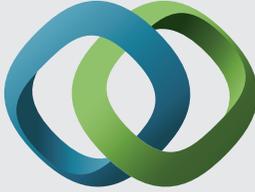
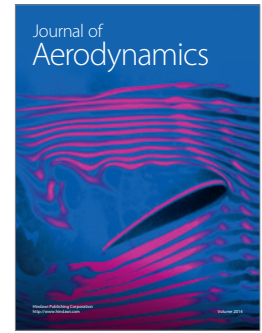
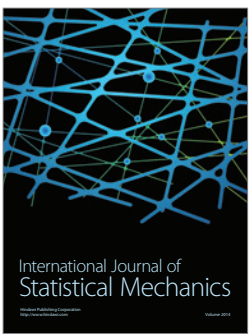
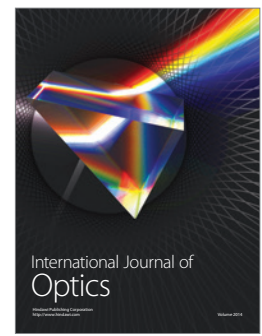
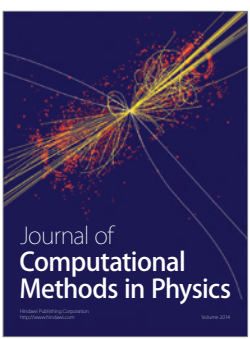
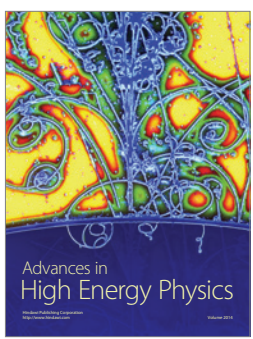
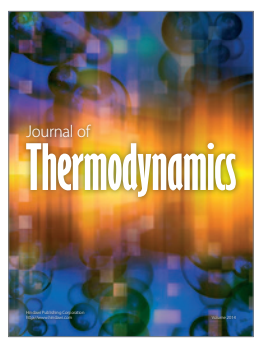
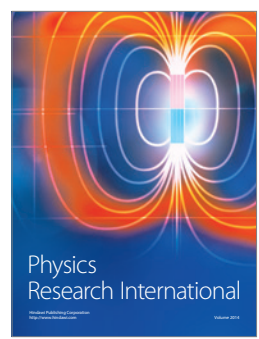
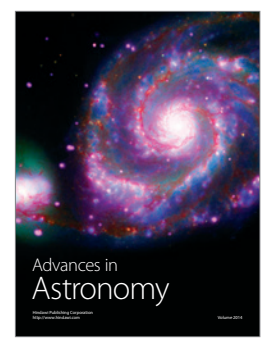
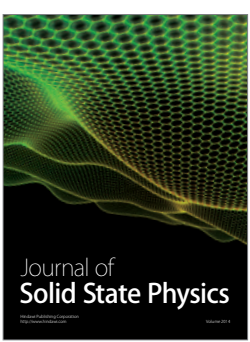
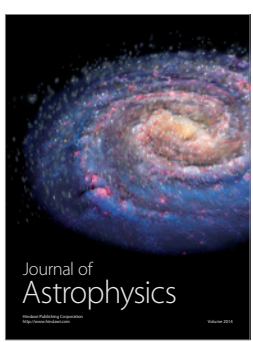
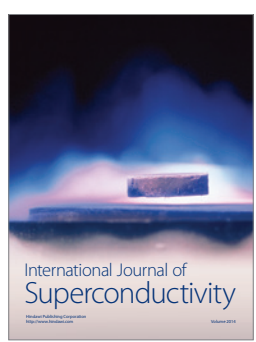
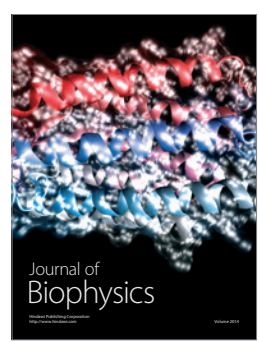
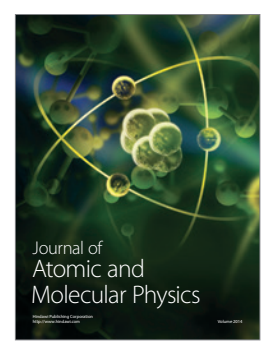